\documentclass[fleqn,usenatbib]{mnras}

\usepackage{newtxtext,newtxmath}

\usepackage[T1]{fontenc}

\DeclareRobustCommand{\VAN}[3]{#2}
\let\VANthebibliography\thebibliography
\def\thebibliography{\DeclareRobustCommand{\VAN}[3]{##3}\VANthebibliography}

\usepackage{graphicx}	
\usepackage{amsmath}	
\usepackage{physics}
\usepackage{subfigure}
\usepackage{chemformula}

\title[Temperature dependent constraint]{Effect of temperature on measurement of fundamental constants using white dwarfs in {\it Gaia}-EDR3 survey}

\author[Uniyal, Kalita and Chakrabarti]{
Akhil Uniyal,$^{1}$\thanks{E-mail: akhil\_uniyal@sjtu.edu.cn}
Surajit Kalita,$^{2}$\thanks{E-mail: surajit.kalita@uct.ac.za; corresponding author} and
Sayan Chakrabarti$^{3}$\thanks{E-mail: sayan.chakrabarti@iitg.ac.in}
\\
$^{1}$Tsung-Dao Lee Institute, Shanghai Jiao Tong University, Shengrong Road 520, Shanghai, 201210, People’s Republic of China\\
$^{2}$High Energy Physics, Cosmology and Astrophysics Theory (HEPCAT) Group, \\ Department of Mathematics and Applied Mathematics, University of Cape Town, Cape Town 7700, South Africa\\
$^{3}$Department of Physics, Indian Institute of Technology, Guwahati 781039, India
}

\date{Accepted XXX. Received YYY; in original form ZZZ}

\pubyear{2023}

\begin{document}
\label{firstpage}
\pagerange{\pageref{firstpage}--\pageref{lastpage}}
\maketitle

\begin{abstract}
Fundamental constants are crucial for comprehending physical mechanisms, but their measurements contain uncertainties due to experimental limitations. We investigate the impact of system temperature on these uncertainties using nearby white dwarfs observed in the {\it Gaia} Early Data Release 3~(EDR3) survey. Using the structures of these white dwarfs, we show that the variation in system temperature can affect the accuracy of measurements for fundamental parameters such as the fine-structure constant and the proton-to-electron mass ratio. This exploration emphasizes the importance of considering the energy of a system while putting bounds on the values of fundamental constants.
\end{abstract}




\section{Introduction}
Testing fundamental couplings that lead to the investigation of the universality of physical laws always remains one of the most interesting fields for researchers~\citep{Uzan:2010pm, Martins:2014iaa}. The spatial variation of the fine-structure constant was investigated using the high-resolution spectroscopic studies of absorption systems along the line of sight of bright quasars~\citep{Webb:2010hc}, and consequently, detailed further tests on this claim were discussed by~\cite{Evans:2014yva} and~\cite{Kotus:2016xxb} using different quasars. Similar tests for compact objects, such as solar-type main-sequence stars~\citep{Adams:2008ad, Vieira:2012pu}, population III stars~\citep{Coc:2010fvl}, and neutron stars~\citep{Perez-Garcia:2012wja} were also carried out to put bounds on the fine-structure constant, the Newton gravitational constant, etc. It is worth mentioning that due to significant uncertainties associated with the nuclear processes, these tests are not as strong as the quasar tests. Moreover, as these two tests were carried out in different physical environments, they led to completely independent results. Nevertheless, the possible dependencies of fundamental couplings on the local environment, such as the strength of the local gravitational field in the presence of compact objects, should also be handled carefully.

More recently, researchers put constraints on the fine structure constant $\alpha$ as well as the proton-to-electron mass ratio $\mu$ using white dwarf (WD) observations from different surveys~\citep{2017PhRvD..96h3012M,Kalita:2023hcl}. WDs are the end state of main-sequence stars possessing mass less than approximately $(10\pm 2)\rm\,M_\odot$~\citep{2018MNRAS.480.1547L}. Earlier, \cite{1935MNRAS..95..207C} showed that a non-rotating and non-magnetized WD can contain a maximum mass of $1.4\rm\,M_\odot$; famously known as the Chandrasekhar mass-limit. In order to describe the phenomenology of a WD, one needs to understand its underlying equation of state (EoS) which describes the state of matter within it.

So far in the literature, there exist three major EoSs which describe the degenerate matter of the WDs known as the Chandrasekhar EoS~\citep{1931ApJ....74...81C}, the Salpeter EoS~\citep{1961ApJ...134..669S, 1969ApJ...155..183S}, and the relativistic Feynman-Metropolis-Teller (RFMT) EoS~\citep{1949PhRv...75.1561F, 2011PhRvC..83d5805R}. Among them, the RFMT EoS takes care of the coulomb interactions and the local inhomogeneities of the relativistic electrons making it a more generalized form of the Chandrasekhar and Salpeter EoSs. The contributions from these terms decrease WD masses but enhance their radii as compared to those obtained from the Chandrasekhar and Salpeter EoSs. A detailed comparison of these EoSs was studied by~\cite{Rotondo:2011zz} including their advantages and drawbacks. The widely used polytropic EoSs for WDs are the non-relativistic and relativistic limiting cases of the Chandrasekhar and Salpeter EoSs~\citep{YaBZel'dovich_1966, 1983bhwd.book.....S}. It is worth mentioning that RFMT EoS shows significant deviation from the other EoSs only at relatively low densities, primarily below $10^4\rm\,g\,cm^{-3}$~\citep{deCarvalho:2013rea}. In general, the observed WDs cannot possess such low core densities, which is responsible for their overall masses, and hence it does not significantly affect the overall structure of the observed WDs. Thus, we consider the Chandrasekhar EoS and its possible modification in the presence of temperature in this work. Our target is to incorporate the temperature effect into the EoS so that the mass--radius relation can be matched with the observed WD data and thereby look for the effects of temperature on the constraints of $\alpha$ and $\mu$.

Previously, various observations and experiments were performed to put different constraints on $\alpha$ and $\mu$. Using molecular hydrogen transitions in the quasar Q$0528-250$ spectra, \cite{king2011new} estimated $\Delta \mu /\mu = (0.3 \pm 5.1)\times 10^{-6}$ at a redshift $z=2.811$. An improved constraint of $\Delta \mu /\mu = (0.0 \pm 1.0) \times 10^{-7}$ was found by analysing \ch{NH3} from PKS\,$1830-211$ quasar spectrum at $z=0.89$~\citep{2014PhRvL.113l3002B}. However, this bound comes with the limitation that it is only valid for low redshift systems with $z \le 1.0$~\citep{Rahmani:2012ze}. At higher redshifts, using 27 Ritz wavelengths of [Fe II] lines of quasar spectra, it was later found that $\Delta \mu /\mu < 10^{-5}$ at $2 \le z \le 3$~\citep{Le:2019ijj}. Observations of 21-cm and molecular hydrogen absorption spectra from higher redshift systems such as J$1337+3152$ at $z \approx 3.17$ slightly loosen this bound with $\Delta \mu /\mu = (-1.7 \pm 1.7) \times 10^{-6}$~\citep{Srianand:2010un}. Eventually, researchers proposed some other bounds on these fundamental parameters using cosmological data~\citep{2015PhRvC..92a4319D,2015Ap&SS.357....4K,2017PhRvC..96d5802M,2018MNRAS.474.1850H}; a detailed comprehensive review was produced by~\cite{2017RPPh...80l6902M}. Apart from the cosmological bounds, researchers also obtained similar bounds using astronomical objects such as WDs. Lyman transitions of \ch{H2} in the spectra of WDs GD\,133 and G$29-38$ yield the bounds $\Delta \mu /\mu = (-2.7 \pm 4.9) \times 10^{-5}$ and $\Delta \mu /\mu = (-5.8 \pm 4.1) \times 10^{-5}$, respectively~\citep{2014PhRvL.113l3002B}, whereas the analysis of Fe\,V spectra from another WD G191$-$B2B shows $\Delta \alpha / \alpha = (6.36\pm2.27)\times10^{-5}$~\citep{2021MNRAS.500.1466H}. Note that spectroscopic measurement of $\alpha$ and $\mu$ can be made on the surface of the WDs~\citep{Berengut:2013dta,2014PhRvL.113l3002B}, and it can be considered the most promising way to put bounds on some fundamental constants.

Moreover, it was shown that the standard WD mass--radius relation gets affected by the spacetime variation of fundamental couplings. Using the masses and radii from a simulated catalog of 100 WDs, \cite{2017PhRvD..96h3012M} found the bound to be $\Delta \alpha / \alpha = (2.7 \pm 9.1) \times 10^{-5}$. It is worth mentioning that these 100 WDs are not the actual WD data; they are instead simulated ones within the mass range of $0.3\,\textup{M}_\odot < M < 1.2\,\textup{M}_\odot$ using the standard Chandrasekhar mass--radius relation. The actual masses and radii of WDs indeed deviate from this relation as confirmed by different astronomical surveys, such as Sloan Digital Sky Survey (SDSS), {\it Gaia}, etc. In our previous work, with the help of massive WDs observed in {\it Gaia} Data Release 2 survey, we obtained bounds on $\alpha$ and $\mu$ under the Newtonian gravity as well as in modified gravity theory and thereby showed that these bounds are significantly affected by the underlying gravity theories~\citep{Kalita:2023hcl}. Through our exploration, it was firmly established that modified gravity has a significant impact on dense astrophysical scenarios. This has also led to the revelation of stronger bounds of fundamental parameters under alternate gravity theories. 

In this paper, we consider the grand unified theory models for the variations of relevant couplings depending on the two dimensionless parameters $\mathsf{R}$ and $\mathsf{S}$~\citep{Campbell:1994bf,Coc:2006sx}. Using WDs from {\it Gaia} Early Data Release 3~(EDR3) catalogue~\citep{2021MNRAS.508.3877G}, we show that the variation in temperature can significantly affect the constraints on $\alpha$ and $\mu$. This article is organized as follows. In Section~\ref{Sec2}, we first obtain the temperature-dependent EoS for WD matter and thereby the relations for the bounds of $\alpha$ and $\mu$. In Section~\ref{Sec3}, we describe our selection of dataset using {\it Gaia}-EDR3 survey, which we use to constrain $\alpha$ and $\mu$ for different temperatures. Finally, Section~\ref{Sec4} discusses these results and provides some parting thoughts as conclusions in Section~\ref{Sec5}.

\section{Effect of temperature on the white dwarf structure}\label{Sec2}

As we are going to study the WD structures in the presence of temperature, let us first discuss the corresponding EoS, which can be determined through the pressure and energy density of the matter present inside. Generally, except the envelope region near the surface (which is comparatively much smaller than the total radius of WD), the pressure $P$ inside a WD is dominated by the electron gas pressure $P_\text{e}$; the pressure due to the positively charged nuclei $P_\text{N}$ is insignificant. On the other hand, the energy density $\mathcal{E}$ is dominated by that of the nuclei $\mathcal{E}_\text{N}$ and the energy density of the electrons $\mathcal{E}_\text{e}$ is negligibly small. These approximations were considered by~\cite{1931ApJ....74...81C} and mathematically they can be represented as
\begin{align}
    P &= P_\text{N}+P_\text{e} \approx P_\text{e}
\end{align}
and
\begin{align}
    \mathcal{E} &= \mathcal{E}_\text{N}+\mathcal{E}_\text{e} \approx \mathcal{E}_\text{N} = \mu_\text{e} m_\text{u} c^2 n_\text{e},
\end{align}
where $m_\text{u}=1.6604 \cross 10^{-24}\rm\,g$ is the atomic mass unit, $c$ is the speed of light, $n_\text{e}$ is the electron number density, and $\mu_\text{e}$ is the mean molecular weight per electron. Because electrons are fermions, they follow the Fermi-Dirac statistics, and the electron number density is given by~\citep{boshkayev2016equilibrium,2018ARep...62..847B}
\begin{align}
    n_\text{e} &= \frac{2}{h^3}\int_0^\infty \frac{4\pi p^2 \dd{p}}{\exp[\frac{\tilde{E}(p)-\tilde{\mu}_\text{e}}{k_\text{B} T}]+1} \\
    &= \frac{8\pi \sqrt{2}}{h^3} m_\text{e}^3 c^3 \beta^{3/2} \left[F_{1/2}(\eta, \beta) + \beta F_{3/2}(\eta, \beta)\right],
\end{align}
where $k_\text{B}$ is the Boltzmann constant, $h$ is the Planck constant, $m_\textbf{{e}}$ is the rest mass of an electron, $\tilde{\mu}_\text{e}$ is the chemical potential, $T$ is the temperature, $\beta=k_\text{B}T/(m_\text{e} c^2)$, $p$ is the momentum of the electron, $\tilde{E}(p)= \sqrt{p^2c^2+m_\text{e}^2 c^4} - m_\text{e}c^2$ is the kinetic energy, and 
\begin{equation}
F_{k} (\eta,\beta)=\int_{0} ^{\infty} \frac{t^k \sqrt{1+(\beta/2)t}}{1+ e^{t-\eta}}\dd{t},
\label{eq:Fk1}
\end{equation}
with $t=\tilde{E}(p)/(k_\text{B}T)$ being a dimensionless parameter and $\eta=\tilde{\mu}_\text{e}/(k_\text{B} T)$ the normalised chemical potential energy~\citep{1999ApJS..125..277T, de2014relativistic}. Consequently, the total pressure $P$ and matter density $\rho$ are respectively given by~\citep{de2014relativistic}
\begin{align}\label{pe}
P &= \frac{16\pi \sqrt{2}}{h^3} m_\text{e}^4 c^5 \beta^{5/2} \left[ F_{3/2} (\eta,\beta)+ \frac{\beta}{2} F_{5/2} (\eta,\beta) \right]
\end{align}
and
\begin{align}\label{rhoe}
\rho &= \mu_\text{e} m_\text{p} n_\text{e} = \frac{8\pi \sqrt{2}}{h^3} \mu_\text{e} m_\text{p} m_\text{e}^3 c^3 \beta^{3/2} \left[F_{1/2}(\eta, \beta) + \beta F_{3/2}(\eta, \beta)\right].
\end{align}
It is evident from these equations that the EoS depends on the temperature of the matter. The complexity of the function $F_k(\eta,\beta)$ does not allow us to solve these two equations analytically, and therefore we solve them numerically. Using this EoS, we need to solve the pressure balance and mass estimate equations (together known as hydrostatic balance equations) to obtain the structures of WDs at different temperatures. Due to the large size of observed WDs in our data sample, we can study them within the Newtonian framework. The hydrostatic balance equations are given by
\begin{align}
    \dv{P(r)}{r} &= -\frac{G M(r) \rho(r)}{r^2}
\end{align}
and
\begin{align}
    \dv{M(r)}{r} &= 4 \pi r^2 \rho(r),
\end{align}
where $G$ is the Newton gravitational constant and $M(r)$ is the mass of the WD within a radius $r$. Since we consider the effect of temperature, one, in principle, needs to simultaneously solve the radiative energy transport and the energy conservation equations, which are respectively given by
\begin{align}
    \dv{T(r)}{r} &= -\frac{3 L(r) \kappa(r) \rho(r)}{64\pi r^2 \sigma T(r)^3}
\end{align}
and
\begin{align}
    \dv{L(r)}{r} &= 4 \pi r^2 \rho(r) \epsilon(r),
\end{align}
where $L$ is the luminosity, $\epsilon$ is the power produced per unit mass of WD matter, $\kappa$ is the opacity, and $\sigma$ is the Stefan-Boltzmann constant. We can divide a WD into two parts: the core and the envelope. In the core, electron degeneracy pressure dominates over ideal gas pressure due to high density, whereas in the envelope region, it is the opposite. Envelopes are typically only a few kilometers in radius, much smaller compared to the WD radius. Additionally, most of the WD mass accumulates in the core. Previous studies suggest that temperature does not vary significantly in the core region and drops steeply in the envelope~\citep{2022ApJ...925..133B}. Hence, we can safely ignore the $\dv*{T}{r}$ and $\dv*{L}{r}$ equations, as they have negligible effects on the WD mass and radius. We keep the temperature contribution only through the WD EoS.

Our interest lies in finding out the bounds on $\alpha$ and $\mu$. Considering that the Planck mass is fixed but particle mass and quantum chromodynamics (QCD) scale can vary,~\citep{Coc:2006sx} provided the following uncertainties in electron and proton masses
\begin{align}\label{e2.11}
    \frac{\Delta m_\text{e}}{m_\text{e}} &= \frac{1}{2}\left(1+\mathsf{S}\right)\frac{\Delta \alpha}{\alpha}
\end{align}
and
\begin{align}\label{e2.12}
    \frac{\Delta m_\text{p}}{m_\text{p}} &= \left[\frac{4}{5}\mathsf{R} +\frac{1}{5} \left(1+\mathsf{S}\right)\right]\frac{\Delta \alpha}{\alpha},
\end{align}
where $\mathsf{R}$ and $\mathsf{S}$ are dimensionless phenomenological parameters. Thereby the uncertainty in $\mu$ can be written as
\begin{equation}\label{e2.13}
    \frac{\Delta \mu}{\mu} = \left[\frac{4}{5}\mathsf{R}-\frac{3}{10}\left(1+\mathsf{S}\right)\right]\frac{\Delta \alpha}{\alpha}.
\end{equation}
The values of $\mathsf{R}$ and $\mathsf{S}$ tend to vary based on different observations. For instance, the data from the Wilkinson Microwave Anisotropy Probe (WMAP) suggest that $\mathsf{R} \approx 36$ and $\mathsf{S} \approx 160$~\citep{Coc:2006sx}, whereas a dilaton-type model gives $\mathsf{R} \approx 109$ and $\mathsf{S} \approx 0$~\citep{nakashima2010constraining}. However, in this paper, we consider values obtained from the astrophysical observations of a BL Lac object PKS\,$1413+135$, which suggest $\mathsf{R}=278\pm24$ and $\mathsf{S}=742\pm65$~\citep{2014MmSAI..85..113M}. We now insert these uncertainties of electron and proton masses in Equations~\eqref{pe} and~\eqref{rhoe} to further solve the hydrostatic balance equations for a fixed temperature and different $\Delta \alpha/\alpha$.

\section{Effect of temperature on the constraint of fine-structure constant}\label{Sec3}

In this work, we analyze the effect of temperature on the bounds of $\alpha$ and $\mu$ using {\it Gaia}-EDR3 data reported by~\cite{2021MNRAS.508.3877G}\footnote{\url{https://warwick.ac.uk/fac/sci/physics/research/astro/research/catalogues}}. The dataset contains 1\,280\,266 objects, each associated with a probability of being a WD ($P_\text{WD}$) and geometric distance. Among them, masses are measured for 298\,317 objects considering that they are comprised of pure hydrogen atmosphere. In this work, we limit our analysis to objects within a geometric distance of 50\,pc and which are almost certain to be WD candidates with $P_\text{WD}\geq 0.9$. This reduces our sample size to 2094. We further use their masses and $\log g$ values with $g$ being the surface gravity to obtain the radii.

\begin{figure}
	\centering
	\includegraphics[scale=0.5]{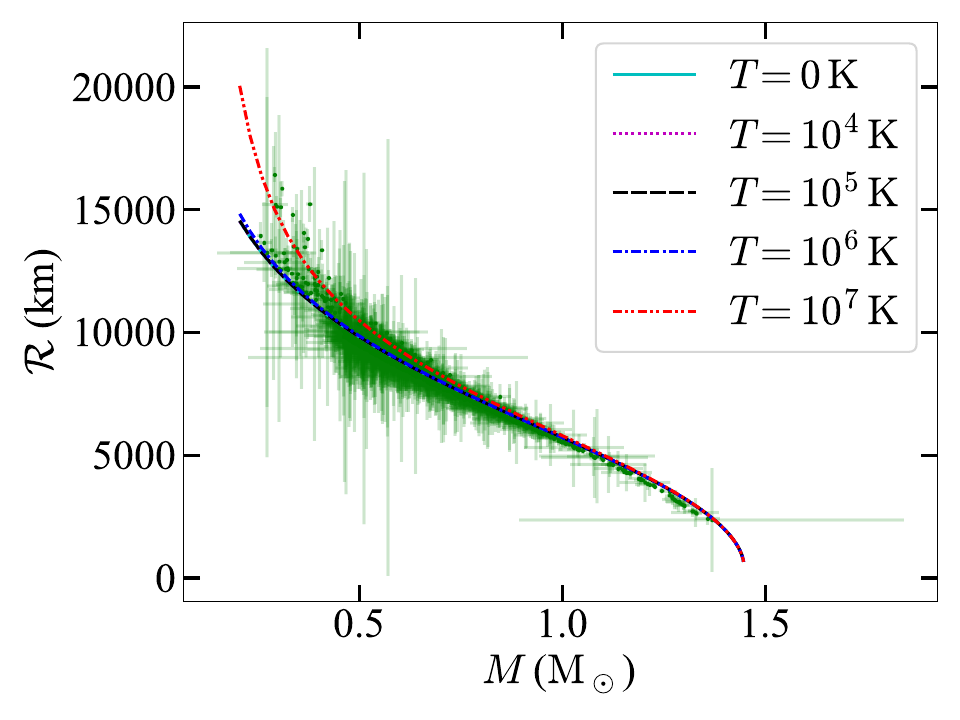}
	\caption{The green scattered points are masses and radii of WDs with their respective error bars obtained from the {\it Gaia}-EDR3 survey, which are within a radius of $50\rm\,pc$. Different coloured lines depict WDs' theoretical mass--radius curves for different temperatures assuming $\mu_\text{e}=2$.}
	\label{Fig: Gaia}
\end{figure}
Figure~\ref{Fig: Gaia} shows the masses and radii of these WDs along with their respective error bars. Different lines in that plot show the theoretical mass--radius curves obtained by solving the hydrostatic balance equations together with Equations~\eqref{pe} and~\eqref{rhoe} for different temperatures. We assume $\mu_\text{e}=2$ Throughout this work indicating the cores of the WDs are comprised of carbon or oxygen or any other similar matter. It is evident that the effect of temperature is more prominent for less massive WDs and this significance gradually drops with the increase in mass. It is also noticeable from this figure that the masses and radii are more scattered towards the low mass region whereas they nearly follow the Chandrasekhar zero-temperature mass--radius curve as mass increases. The temperature of a WD is unequivocally influenced by its evolutionary history and surrounding environment. Because temperature exerts a greater impact on low densities, it leads to more sporadic data points toward the low mass regimes in the mass--radius plot. It is important to note that there are other physical effects such as rotation and magnetic fields which may potentially impact the shape and size of WDs. However, the data points in the high mass regimes suggest that they do not have discernible effects on the mass and size of WDs. As a result, we do not factor in these effects when deriving theoretical mass-radius curves for WDs.

\begin{figure}
	\centering
	\subfigure[Bounds on $\alpha$]{\includegraphics[scale=0.5]{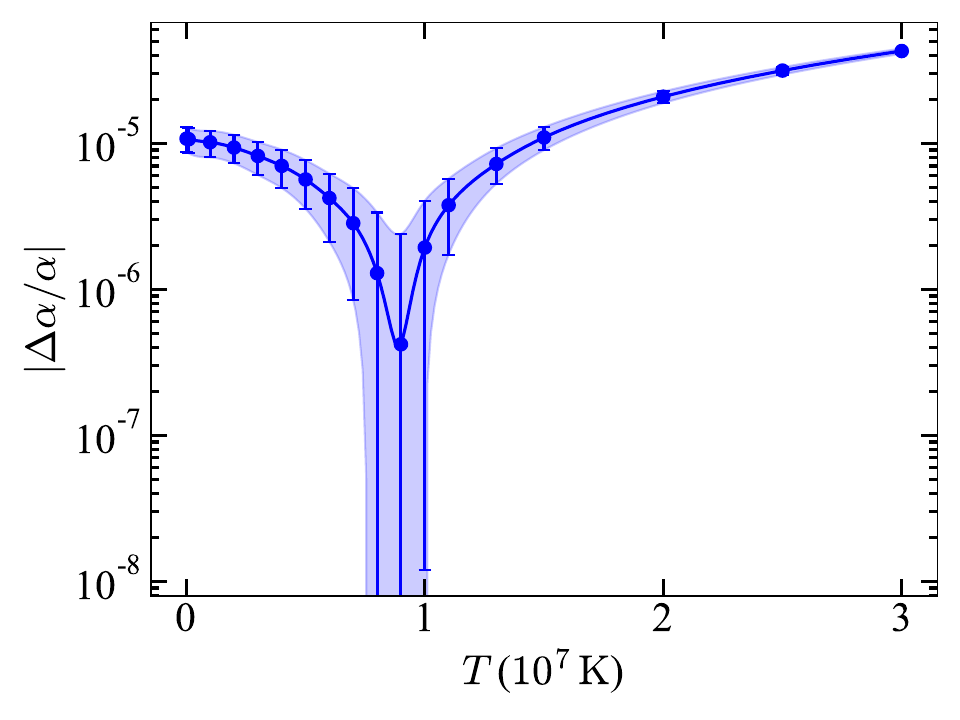}}
	\subfigure[Bounds on $\mu$]{\includegraphics[scale=0.5]{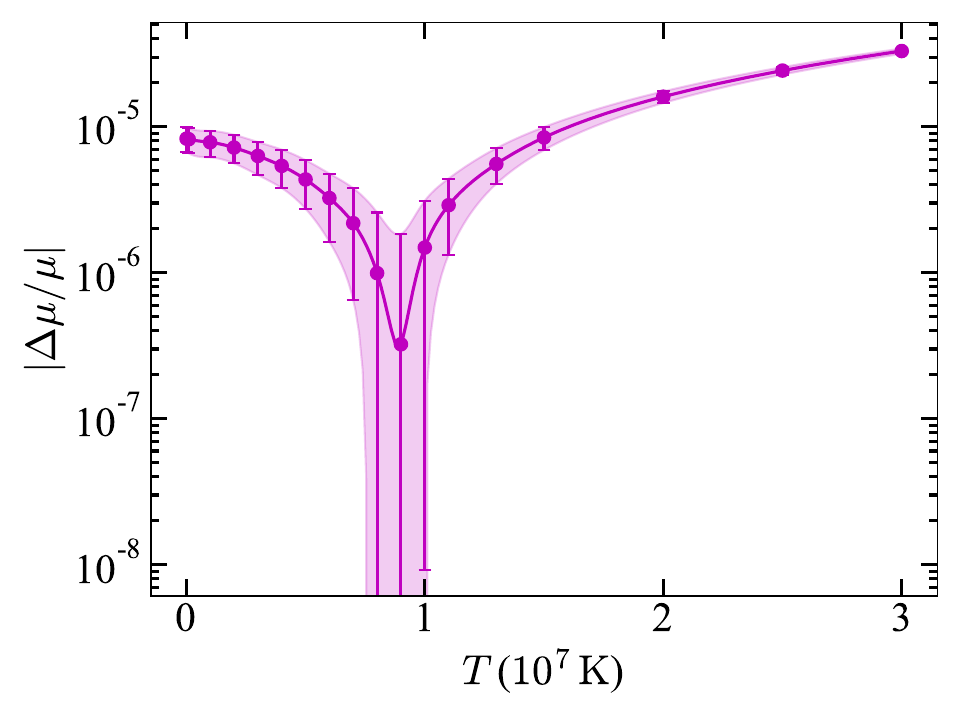}}
	\caption{Variation of uncertainties of fundamental constants as a function of temperature along with their corresponding 3-$\sigma$ error bars. Note that the sizes of error bars might look different as the vertical axis is plotted in logarithmic scale.}
	\label{Fig: constraint}
\end{figure}
We need to understand the goodness of the theoretical mass--radius curves over the data points, which eventually helps in estimating the quantity $\Delta\alpha/\alpha$. In order to do so, one may define the following quantity~\citep{2016PhRvL.116o1103J}
\begin{align}
    \chi^2 = \sum_{i=1}^N \frac{(M-M_i)^2}{\sigma_{M_i}^2}+\frac{(\mathcal{R}_\text{Th}(M)-\mathcal{R}_i)^2}{\sigma_{\mathcal{R}_i}^2},
\end{align}
which needs to be minimized over $\Delta\alpha/\alpha$. Here $M_i$ and $\mathcal{R}_i$ are respectively the observed mass and radius of each data point with $\sigma_{M_i}$ and $\sigma_{\mathcal{R}_i}$ being their respective error bars, whereas $\mathcal{R}_\text{Th} (M)$ denotes the theoretical radius for mass $M$. However, in our case, the radii of these WDs are not independently measured from observations; we rather obtain them using the measured values of $M$ and $\log g$. Thus, the statistical method demands that we are supposed to consider only one term in the aforementioned $\chi^2$ function. For convenience, we define the $\chi^2$ function as
\begin{align}\label{Eq: chi square}
    \chi^2 = \sum_{i=1}^N \frac{(\mathcal{R}_\text{Th}(M)-\mathcal{R}_i)^2}{\sigma_{\mathcal{R}_i}^2}.
\end{align}
We now minimize this function for different $\Delta\alpha/\alpha$ over each temperature. Figure~\ref{Fig: constraint}(a) shows $\abs{\Delta\alpha/\alpha}$ as well as 3-$\sigma$ confidence intervals for which $\chi^2$ is minimized at different temperatures. The detailed analysis of the statistical method is discussed in Appendix~\ref{Appendix}. It is evident that until the temperature significantly affects the WD mass--radius curves, $\abs{\Delta\alpha/\alpha}$ nearly remains constant. However, above $10^6\rm\, K$ it starts decreasing and reaches a minimum approximately at $9\times 10^6\rm\, K$. A further increase in temperature increases the $\abs{\Delta\alpha/\alpha}$ indefinitely because such high temperatures are not possible inside of a WD. Substituting these values of $\abs{\Delta\alpha/\alpha}$ in Equation~\eqref{e2.13}, we calculate $\abs{\Delta\mu/\mu}$ and their corresponding 3-$\sigma$ error bars. which is depicted in Figure~\ref{Fig: constraint}(b) to observe that the variation of $\abs{\Delta\mu/\mu}$ with respect to temperature behaves the same way as $\abs{\Delta\alpha/\alpha}$.

\section{Discussion}\label{Sec4}

This work delves into the impact of system temperature on the measurement of fundamental constants $\alpha$ and $\mu$. Our methodology involves selecting nearby WDs from the {\it Gaia}-EDR3 survey, which are within a distance of $50\rm\,pc$ and with the probability of being a WD greater than or equal to $0.9$. After comparing their structures with the theoretical mass--radius curves for different temperatures, we have observed that the effect of temperature is prominent for the less massive WDs. With the increase in WD mass, the influence of temperature on the overall structure of WDs gradually diminishes, and above $1\rm\,M_\odot$, it is almost insignificant. Since the uncertainties in electron and proton mass can be related to the uncertainties in $\alpha$, we have used these mass uncertainties in the finite temperature EoS assuming that the Planck mass is fixed but particle mass and QCD scale can vary~\citep{Coc:2006sx}. Through our analysis, we have determined the optimal value of $\Delta \alpha / \alpha$ for each temperature by minimizing the $\chi^2$ function. By repeating this process for various temperatures, we have established the bounds on $\alpha$ and $\mu$. These values serve as the most stringent constraints on $\alpha$ and $\mu$ at their respective temperatures for these WD data. Our findings reveal that the constraints on these parameters become tighter as the temperature rises, culminating the tightest constraints of $\abs{\Delta\alpha/\alpha} = 4.203\times10^{-7}$ and $\abs{\Delta\mu/\mu} = 3.233\times10^{-7}$ at approximately $9\times10^6\rm\,K$ before they relax indefinitely.

We have made some justifiable assumptions in our work. Specifically, in Section~\ref{Sec2}, we have excluded $\dv*{T}{r}$ and $\dv*{L}{r}$ equations as they do not significantly impact the masses and radii of WDs. We have solely focused on the temperature contribution through the WD EoS. Note that we have only considered the nearby WDs. It is unlikely that choosing all WDs would significantly affect the final results of Figure~\ref{Fig: constraint}. This is because the mass distribution of the selected WDs is similar to that of the complete sample of WDs. Moreover, the influence of magnetic fields and rotation, which can increase the mass of a WD, have not been taken into account in our study. The magnetic-to-gravitational~(ME/GE) and kinetic-to-gravitational~(KE/GE) energy ratios would need to be exceptionally high for these factors to significantly impact the WD's mass, potentially surpassing the proposed limits by~\cite{1989MNRAS.237..355K} and~\cite{2009MNRAS.397..763B}. Therefore, the inferred masses and radii of WDs in the {\it Gaia}-EDR3 catalog would remain unchanged unless the ME/GE and KE/GE ratios are exceedingly high. Furthermore, we have chosen $\mu_\text{e}=2$ which indicates the presence of carbon, oxygen, neon, or other similar elements at the core of these WDs. This assumption is supported by Figure~\ref{Fig: Gaia} where all the massive WDs nearly follow the mass--radius relation with $\mu_\text{e}=2$. Nevertheless, we do not rule out the possibility of the presence of other elements although they would not significantly alter the final outcome.

\section{Conclusions}\label{Sec5}
The significance of system temperature in determining fundamental constants and couplings is emphasized in this work. In the introduction, we have already mentioned some bounds on $\alpha$ and $\mu$ from different observations. Note that these bounds are different while considering objects at different redshifts. The thermal history of the universe shows that the temperature gradually decreases after the Big Bang, and hence different redshifts indicate the different temperatures of our universe. In this way, these aforementioned results also indirectly indicate that the constraints on fundamental parameters are affected by the temperature. 

In our work, by using WD data from an astronomical survey, we have explicitly shown that these constraints can vary as the temperature alters. Note that this statement is true in a broader sense, although the situation can be different for spectroscopic analyses. For instance, in the case of quasar absorption measurements of $\alpha$, which are discussed in the Introduction, it is true that the gas temperature is relevant and it can be treated as a free parameter of equal status to all other free parameters for which they are solved. However, if the quasar analysis is restricted to the some atomic species, e.g. [Fe~II], the measurement is independent of temperature. It is to be noted that at a certain temperature, two different observations might yield two different constraints for the same parameter. In this scenario, one must consider the tightest one at that particular point. Our work stands out for providing some of the tightest constraints on $\alpha$ and $\mu$ compared to existing works in literature, which are mentioned in the Introduction. This exploration highlights that the significance of taking into account the system energy when defining limits for fundamental constants is undeniable. As we move forward, conducting further analysis through physical observations can unlock a deeper understanding of these outcomes.

\section*{Acknowledgements}

We thank the reviewer for their constructive suggestions to improve the quality of the manuscript, especially the statistical analysis section. A.U. thanks K. Boshkayev of Al-Farabi Kazakh National University for their helpful suggestion to understand the temperature-dependent equation of state during the compilation of this work. S.K. would like to acknowledge support from the South African Research Chairs Initiative~(SARChI) of the Department of Science and Technology~(DST) and the National Research Foundation~(NRF). The work of S.C. is supported by Mathematical Research Impact Centric Support (MATRICS) from the Science and Engineering Research Board (SERB) of India through grant MTR/2022/000318. Computations were performed using facilities provided by the University of Cape Town’s ICTS High Performance Computing team: \href{https://ucthpc.uct.ac.za/}{hpc.uct.ac.za}.

\section*{Data availability}
The data underlying this article were accessed from \url{https://warwick.ac.uk/fac/sci/physics/research/astro/research/catalogues}. The derived data generated in this research will be shared on reasonable request to the corresponding author.



\bibliographystyle{mnras}
\bibliography{bibliography} 

\begin{thebibliography}{}
\makeatletter
\relax
\def\mn@urlcharsother{\let\do\@makeother \do\$\do\&\do\#\do\^\do\_\do\%\do\~}
\def\mn@doi{\begingroup\mn@urlcharsother \@ifnextchar [ {\mn@doi@}
  {\mn@doi@[]}}
\def\mn@doi@[#1]#2{\def\@tempa{#1}\ifx\@tempa\@empty \href
  {http://dx.doi.org/#2} {doi:#2}\else \href {http://dx.doi.org/#2} {#1}\fi
  \endgroup}
\def\mn@eprint#1#2{\mn@eprint@#1:#2::\@nil}
\def\mn@eprint@arXiv#1{\href {http://arxiv.org/abs/#1} {{\tt arXiv:#1}}}
\def\mn@eprint@dblp#1{\href {http://dblp.uni-trier.de/rec/bibtex/#1.xml}
  {dblp:#1}}
\def\mn@eprint@#1:#2:#3:#4\@nil{\def\@tempa {#1}\def\@tempb {#2}\def\@tempc
  {#3}\ifx \@tempc \@empty \let \@tempc \@tempb \let \@tempb \@tempa \fi \ifx
  \@tempb \@empty \def\@tempb {arXiv}\fi \@ifundefined
  {mn@eprint@\@tempb}{\@tempb:\@tempc}{\expandafter \expandafter \csname
  mn@eprint@\@tempb\endcsname \expandafter{\@tempc}}}

\bibitem[\protect\citeauthoryear{{Adams}}{{Adams}}{2008}]{Adams:2008ad}
{Adams} F.~C.,  2008, \mn@doi [\jcap] {10.1088/1475-7516/2008/08/010}, \href
  {https://ui.adsabs.harvard.edu/abs/2008JCAP...08..010A} {2008, 010}

\bibitem[\protect\citeauthoryear{{Bagdonaite}, {Salumbides}, {Preval},
  {Barstow}, {Barrow}, {Murphy}  \& {Ubachs}}{{Bagdonaite}
  et~al.}{2014}]{2014PhRvL.113l3002B}
{Bagdonaite} J.,  {Salumbides} E.~J.,  {Preval} S.~P.,  {Barstow} M.~A.,
  {Barrow} J.~D.,  {Murphy} M.~T.,   {Ubachs} W.,  2014, \mn@doi [\prl]
  {10.1103/PhysRevLett.113.123002}, \href
  {https://ui.adsabs.harvard.edu/abs/2014PhRvL.113l3002B} {113, 123002}

\bibitem[\protect\citeauthoryear{{Berengut}, {Flambaum}, {Ong}, {Webb},
  {Barrow}, {Barstow}, {Preval}  \& {Holberg}}{{Berengut}
  et~al.}{2013}]{Berengut:2013dta}
{Berengut} J.~C.,  {Flambaum} V.~V.,  {Ong} A.,  {Webb} J.~K.,  {Barrow} J.~D.,
   {Barstow} M.~A.,  {Preval} S.~P.,   {Holberg} J.~B.,  2013, \mn@doi [\prl]
  {10.1103/PhysRevLett.111.010801}, \href
  {https://ui.adsabs.harvard.edu/abs/2013PhRvL.111a0801B} {111, 010801}

\bibitem[\protect\citeauthoryear{{Bhattacharya}, {Hackett}, {Gupta}, {Tout}  \&
  {Mukhopadhyay}}{{Bhattacharya} et~al.}{2022}]{2022ApJ...925..133B}
{Bhattacharya} M.,  {Hackett} A.~J.,  {Gupta} A.,  {Tout} C.~A.,
  {Mukhopadhyay} B.,  2022, \mn@doi [\apj] {10.3847/1538-4357/ac450b}, \href
  {https://ui.adsabs.harvard.edu/abs/2022ApJ...925..133B} {925, 133}

\bibitem[\protect\citeauthoryear{{Boshkayev}}{{Boshkayev}}{2018}]{2018ARep...62..847B}
{Boshkayev} K.,  2018, \mn@doi [Astronomy Reports] {10.1134/S106377291812017X},
  \href {https://ui.adsabs.harvard.edu/abs/2018ARep...62..847B} {62, 847}

\bibitem[\protect\citeauthoryear{{Boshkayev}, {Rueda}, {Zhami}, {Kalymova}  \&
  {Balgymbekov}}{{Boshkayev} et~al.}{2016}]{boshkayev2016equilibrium}
{Boshkayev} K.~A.,  {Rueda} J.~A.,  {Zhami} B.~A.,  {Kalymova} Z.~A.,
  {Balgymbekov} G.~S.,  2016, in International Journal of Modern Physics
  Conference Series. p. 1660129 (\mn@eprint {arXiv} {1510.02024}),
  \mn@doi{10.1142/S2010194516601290}

\bibitem[\protect\citeauthoryear{{Braithwaite}}{{Braithwaite}}{2009}]{2009MNRAS.397..763B}
{Braithwaite} J.,  2009, \mn@doi [\mnras] {10.1111/j.1365-2966.2008.14034.x},
  \href {http://adsabs.harvard.edu/abs/2009MNRAS.397..763B} {397, 763}

\bibitem[\protect\citeauthoryear{{Campbell} \& {Olive}}{{Campbell} \&
  {Olive}}{1995}]{Campbell:1994bf}
{Campbell} B.~A.,  {Olive} K.~A.,  1995, \mn@doi [Physics Letters B]
  {10.1016/0370-2693(94)01652-S}, \href
  {https://ui.adsabs.harvard.edu/abs/1995PhLB..345..429C} {345, 429}

\bibitem[\protect\citeauthoryear{{Chandrasekhar}}{{Chandrasekhar}}{1931}]{1931ApJ....74...81C}
{Chandrasekhar} S.,  1931, \mn@doi [\apj] {10.1086/143324}, \href
  {https://ui.adsabs.harvard.edu/abs/1931ApJ....74...81C} {74, 81}

\bibitem[\protect\citeauthoryear{{Chandrasekhar}}{{Chandrasekhar}}{1935}]{1935MNRAS..95..207C}
{Chandrasekhar} S.,  1935, \mn@doi [\mnras] {10.1093/mnras/95.3.207}, \href
  {http://adsabs.harvard.edu/abs/1935MNRAS..95..207C} {95, 207}

\bibitem[\protect\citeauthoryear{{Coc}, {Nunes}, {Olive}, {Uzan}  \&
  {Vangioni}}{{Coc} et~al.}{2007}]{Coc:2006sx}
{Coc} A.,  {Nunes} N.~J.,  {Olive} K.~A.,  {Uzan} J.-P.,   {Vangioni} E.,
  2007, \mn@doi [\prd] {10.1103/PhysRevD.76.023511}, \href
  {https://ui.adsabs.harvard.edu/abs/2007PhRvD..76b3511C} {76, 023511}

\bibitem[\protect\citeauthoryear{{Coc}, {Ekstr{\"o}m}, {Descouvemont},
  {Meynet}, {Uzan}  \& {Vangioni}}{{Coc} et~al.}{2010}]{Coc:2010fvl}
{Coc} A.,  {Ekstr{\"o}m} S.,  {Descouvemont} P.,  {Meynet} G.,  {Uzan} J.~P.,
  {Vangioni} E.,  2010, in {Tanihara} I.,  {Ong} H.~J.,  {Tamii} A.,
  {Kishimoto} T.,  {Kajino} T.,  {Kubono} S.,   {Shima} T.,  eds,  American
  Institute of Physics Conference Series Vol. 1269, 10th International
  Symposium on Origin of Matter and Evolution of Galaxies: OMEG - 2010. pp
  21--26 (\mn@eprint {arXiv} {0911.2420}), \mn@doi{10.1063/1.3485139}

\bibitem[\protect\citeauthoryear{{Davis} \& {Hamdan}}{{Davis} \&
  {Hamdan}}{2015}]{2015PhRvC..92a4319D}
{Davis} E.~D.,  {Hamdan} L.,  2015, \mn@doi [\prc]
  {10.1103/PhysRevC.92.014319}, \href
  {https://ui.adsabs.harvard.edu/abs/2015PhRvC..92a4319D} {92, 014319}

\bibitem[\protect\citeauthoryear{{Evans} et~al.,}{{Evans}
  et~al.}{2014}]{Evans:2014yva}
{Evans} T.~M.,  et~al., 2014, \mn@doi [\mnras] {10.1093/mnras/stu1754}, \href
  {https://ui.adsabs.harvard.edu/abs/2014MNRAS.445..128E} {445, 128}

\bibitem[\protect\citeauthoryear{{Feynman}, {Metropolis}  \&
  {Teller}}{{Feynman} et~al.}{1949}]{1949PhRv...75.1561F}
{Feynman} R.~P.,  {Metropolis} N.,   {Teller} E.,  1949, \mn@doi [Physical
  Review] {10.1103/PhysRev.75.1561}, \href
  {https://ui.adsabs.harvard.edu/abs/1949PhRv...75.1561F} {75, 1561}

\bibitem[\protect\citeauthoryear{{Gentile Fusillo} et~al.,}{{Gentile Fusillo}
  et~al.}{2021}]{2021MNRAS.508.3877G}
{Gentile Fusillo} N.~P.,  et~al., 2021, \mn@doi [\mnras]
  {10.1093/mnras/stab2672}, \href
  {https://ui.adsabs.harvard.edu/abs/2021MNRAS.508.3877G} {508, 3877}

\bibitem[\protect\citeauthoryear{{Hart} \& {Chluba}}{{Hart} \&
  {Chluba}}{2018}]{2018MNRAS.474.1850H}
{Hart} L.,  {Chluba} J.,  2018, \mn@doi [\mnras] {10.1093/mnras/stx2783}, \href
  {https://ui.adsabs.harvard.edu/abs/2018MNRAS.474.1850H} {474, 1850}

\bibitem[\protect\citeauthoryear{{Hu} et~al.,}{{Hu}
  et~al.}{2021}]{2021MNRAS.500.1466H}
{Hu} J.,  et~al., 2021, \mn@doi [\mnras] {10.1093/mnras/staa3066}, \href
  {https://ui.adsabs.harvard.edu/abs/2021MNRAS.500.1466H} {500, 1466}

\bibitem[\protect\citeauthoryear{{Jain}, {Kouvaris}  \& {Nielsen}}{{Jain}
  et~al.}{2016}]{2016PhRvL.116o1103J}
{Jain} R.~K.,  {Kouvaris} C.,   {Nielsen} N.~G.,  2016, \mn@doi [\prl]
  {10.1103/PhysRevLett.116.151103}, \href
  {https://ui.adsabs.harvard.edu/abs/2016PhRvL.116o1103J} {116, 151103}

\bibitem[\protect\citeauthoryear{{Kalita} \& {Uniyal}}{{Kalita} \&
  {Uniyal}}{2023}]{Kalita:2023hcl}
{Kalita} S.,  {Uniyal} A.,  2023, \mn@doi [\apj] {10.3847/1538-4357/accf1c},
  \href {https://ui.adsabs.harvard.edu/abs/2023ApJ...949...62K} {949, 62}

\bibitem[\protect\citeauthoryear{{King}, {Murphy}, {Ubachs}  \& {Webb}}{{King}
  et~al.}{2011}]{king2011new}
{King} J.~A.,  {Murphy} M.~T.,  {Ubachs} W.,   {Webb} J.~K.,  2011, \mn@doi
  [\mnras] {10.1111/j.1365-2966.2011.19460.x}, \href
  {https://ui.adsabs.harvard.edu/abs/2011MNRAS.417.3010K} {417, 3010}

\bibitem[\protect\citeauthoryear{{Komatsu}, {Eriguchi}  \& {Hachisu}}{{Komatsu}
  et~al.}{1989}]{1989MNRAS.237..355K}
{Komatsu} H.,  {Eriguchi} Y.,   {Hachisu} I.,  1989, \mn@doi [\mnras]
  {10.1093/mnras/237.2.355}, \href
  {http://adsabs.harvard.edu/abs/1989MNRAS.237..355K} {237, 355}

\bibitem[\protect\citeauthoryear{{Kotu{\v{s}}}, {Murphy}  \&
  {Carswell}}{{Kotu{\v{s}}} et~al.}{2017}]{Kotus:2016xxb}
{Kotu{\v{s}}} S.~M.,  {Murphy} M.~T.,   {Carswell} R.~F.,  2017, \mn@doi
  [\mnras] {10.1093/mnras/stw2543}, \href
  {https://ui.adsabs.harvard.edu/abs/2017MNRAS.464.3679K} {464, 3679}

\bibitem[\protect\citeauthoryear{{Kraiselburd}, {Landau}, {Negrelli}  \&
  {Garc{\'\i}a-Berro}}{{Kraiselburd} et~al.}{2015}]{2015Ap&SS.357....4K}
{Kraiselburd} L.,  {Landau} S.~J.,  {Negrelli} C.,   {Garc{\'\i}a-Berro} E.,
  2015, \mn@doi [\apss] {10.1007/s10509-015-2325-4}, \href
  {https://ui.adsabs.harvard.edu/abs/2015Ap&SS.357....4K} {357, 4}

\bibitem[\protect\citeauthoryear{{Lauffer}, {Romero}  \& {Kepler}}{{Lauffer}
  et~al.}{2018}]{2018MNRAS.480.1547L}
{Lauffer} G.~R.,  {Romero} A.~D.,   {Kepler} S.~O.,  2018, \mn@doi [\mnras]
  {10.1093/mnras/sty1925}, \href
  {https://ui.adsabs.harvard.edu/abs/2018MNRAS.480.1547L} {480, 1547}

\bibitem[\protect\citeauthoryear{{Le}}{{Le}}{2019}]{Le:2019ijj}
{Le} T.~D.,  2019, \mn@doi [Chinese Journal of Physics]
  {10.1016/j.cjph.2019.10.007}, \href
  {https://ui.adsabs.harvard.edu/abs/2019ChJPh..62..252L} {62, 252}

\bibitem[\protect\citeauthoryear{{Magano}, {Vilas Boas}  \& {Martins}}{{Magano}
  et~al.}{2017}]{2017PhRvD..96h3012M}
{Magano} D.~M.~N.,  {Vilas Boas} J.~M.~A.,   {Martins} C.~J.~A.~P.,  2017,
  \mn@doi [\prd] {10.1103/PhysRevD.96.083012}, \href
  {https://ui.adsabs.harvard.edu/abs/2017PhRvD..96h3012M} {96, 083012}

\bibitem[\protect\citeauthoryear{{Martins}}{{Martins}}{2015}]{Martins:2014iaa}
{Martins} C.~J.~A.~P.,  2015, \mn@doi [General Relativity and Gravitation]
  {10.1007/s10714-014-1843-7}, \href
  {https://ui.adsabs.harvard.edu/abs/2015GReGr..47.1843M} {47, 1843}

\bibitem[\protect\citeauthoryear{{Martins}}{{Martins}}{2017}]{2017RPPh...80l6902M}
{Martins} C.~J.~A.~P.,  2017, \mn@doi [Reports on Progress in Physics]
  {10.1088/1361-6633/aa860e}, \href
  {https://ui.adsabs.harvard.edu/abs/2017RPPh...80l6902M} {80, 126902}

\bibitem[\protect\citeauthoryear{{Monteiro}, {Ferreira}, {Juli{\~a}o}  \&
  {Martins}}{{Monteiro} et~al.}{2014}]{2014MmSAI..85..113M}
{Monteiro} A.~M.~R.~V.~L.,  {Ferreira} M.~C.,  {Juli{\~a}o} M.~D.,   {Martins}
  C.~J.~A.~P.,  2014, \memsai, \href
  {https://ui.adsabs.harvard.edu/abs/2014MmSAI..85..113M} {85, 113}

\bibitem[\protect\citeauthoryear{{Mosquera} \& {Civitarese}}{{Mosquera} \&
  {Civitarese}}{2017}]{2017PhRvC..96d5802M}
{Mosquera} M.~E.,  {Civitarese} O.,  2017, \mn@doi [\prc]
  {10.1103/PhysRevC.96.045802}, \href
  {https://ui.adsabs.harvard.edu/abs/2017PhRvC..96d5802M} {96, 045802}

\bibitem[\protect\citeauthoryear{{Nakashima}, {Ichikawa}, {Nagata}  \&
  {Yokoyama}}{{Nakashima} et~al.}{2010}]{nakashima2010constraining}
{Nakashima} M.,  {Ichikawa} K.,  {Nagata} R.,   {Yokoyama} J.,  2010, \mn@doi
  [\jcap] {10.1088/1475-7516/2010/01/030}, \href
  {https://ui.adsabs.harvard.edu/abs/2010JCAP...01..030N} {2010, 030}

\bibitem[\protect\citeauthoryear{{P{\'e}rez-Garc{\'\i}a} \&
  {Martins}}{{P{\'e}rez-Garc{\'\i}a} \& {Martins}}{2012}]{Perez-Garcia:2012wja}
{P{\'e}rez-Garc{\'\i}a} M.~{\'A}.,  {Martins} C.~J.~A.~P.,  2012, \mn@doi
  [Physics Letters B] {10.1016/j.physletb.2012.10.047}, \href
  {https://ui.adsabs.harvard.edu/abs/2012PhLB..718..241P} {718, 241}

\bibitem[\protect\citeauthoryear{Press, Teukolsky, Vetterling  \&
  Flannery}{Press et~al.}{2007}]{press2007numerical}
Press W.~H.,  Teukolsky S.~A.,  Vetterling W.~T.,   Flannery B.~P.,  2007,
  Numerical Recipes 3rd Edition: The Art of Scientific Computing.
Numerical Recipes: The Art of Scientific Computing, Cambridge University Press,
  \url {https://books.google.co.za/books?id=1aAOdzK3FegC}

\bibitem[\protect\citeauthoryear{{Rahmani}, {Srianand}, {Gupta}, {Petitjean},
  {Noterdaeme}  \& {V{\'a}squez}}{{Rahmani} et~al.}{2012}]{Rahmani:2012ze}
{Rahmani} H.,  {Srianand} R.,  {Gupta} N.,  {Petitjean} P.,  {Noterdaeme} P.,
  {V{\'a}squez} D.~A.,  2012, \mn@doi [\mnras]
  {10.1111/j.1365-2966.2012.21503.x}, \href
  {https://ui.adsabs.harvard.edu/abs/2012MNRAS.425..556R} {425, 556}

\bibitem[\protect\citeauthoryear{{Rotondo}, {Rueda}, {Ruffini}  \&
  {Xue}}{{Rotondo} et~al.}{2011a}]{2011PhRvC..83d5805R}
{Rotondo} M.,  {Rueda} J.~A.,  {Ruffini} R.,   {Xue} S.~S.,  2011a, \mn@doi
  [\prc] {10.1103/PhysRevC.83.045805}, \href
  {https://ui.adsabs.harvard.edu/abs/2011PhRvC..83d5805R} {83, 045805}

\bibitem[\protect\citeauthoryear{{Rotondo}, {Rueda}, {Ruffini}  \&
  {Xue}}{{Rotondo} et~al.}{2011b}]{Rotondo:2011zz}
{Rotondo} M.,  {Rueda} J.~A.,  {Ruffini} R.,   {Xue} S.-S.,  2011b, \mn@doi
  [\prd] {10.1103/PhysRevD.84.084007}, \href
  {https://ui.adsabs.harvard.edu/abs/2011PhRvD..84h4007R} {84, 084007}

\bibitem[\protect\citeauthoryear{{Salpeter}}{{Salpeter}}{1961}]{1961ApJ...134..669S}
{Salpeter} E.~E.,  1961, \mn@doi [\apj] {10.1086/147194}, \href
  {https://ui.adsabs.harvard.edu/abs/1961ApJ...134..669S} {134, 669}

\bibitem[\protect\citeauthoryear{{Salpeter} \& {van Horn}}{{Salpeter} \& {van
  Horn}}{1969}]{1969ApJ...155..183S}
{Salpeter} E.~E.,  {van Horn} H.~M.,  1969, \mn@doi [\apj] {10.1086/149858},
  \href {https://ui.adsabs.harvard.edu/abs/1969ApJ...155..183S} {155, 183}

\bibitem[\protect\citeauthoryear{{Shapiro} \& {Teukolsky}}{{Shapiro} \&
  {Teukolsky}}{1983}]{1983bhwd.book.....S}
{Shapiro} S.~L.,  {Teukolsky} S.~A.,  1983, {Black holes, white dwarfs and
  neutron stars: The physics of compact objects}.
Wiley-VCH, New York, \mn@doi{10.1002/9783527617661}

\bibitem[\protect\citeauthoryear{{Srianand}, {Gupta}, {Petitjean}, {Noterdaeme}
   \& {Ledoux}}{{Srianand} et~al.}{2010}]{Srianand:2010un}
{Srianand} R.,  {Gupta} N.,  {Petitjean} P.,  {Noterdaeme} P.,   {Ledoux} C.,
  2010, \mn@doi [\mnras] {10.1111/j.1365-2966.2010.16574.x}, \href
  {https://ui.adsabs.harvard.edu/abs/2010MNRAS.405.1888S} {405, 1888}

\bibitem[\protect\citeauthoryear{{Timmes} \& {Arnett}}{{Timmes} \&
  {Arnett}}{1999}]{1999ApJS..125..277T}
{Timmes} F.~X.,  {Arnett} D.,  1999, \mn@doi [\apjs] {10.1086/313271}, \href
  {https://ui.adsabs.harvard.edu/abs/1999ApJS..125..277T} {125, 277}

\bibitem[\protect\citeauthoryear{{Uzan}}{{Uzan}}{2011}]{Uzan:2010pm}
{Uzan} J.-P.,  2011, \mn@doi [Living Reviews in Relativity]
  {10.12942/lrr-2011-2}, \href
  {https://ui.adsabs.harvard.edu/abs/2011LRR....14....2U} {14, 2}

\bibitem[\protect\citeauthoryear{{Vieira}, {Martins}  \& {Monteiro}}{{Vieira}
  et~al.}{2012}]{Vieira:2012pu}
{Vieira} J.~P.~P.,  {Martins} C.~J.~A.~P.,   {Monteiro} M.~J.~P.~F.~G.,  2012,
  \mn@doi [\prd] {10.1103/PhysRevD.86.043003}, \href
  {https://ui.adsabs.harvard.edu/abs/2012PhRvD..86d3003V} {86, 043003}

\bibitem[\protect\citeauthoryear{{Webb}, {King}, {Murphy}, {Flambaum},
  {Carswell}  \& {Bainbridge}}{{Webb} et~al.}{2011}]{Webb:2010hc}
{Webb} J.~K.,  {King} J.~A.,  {Murphy} M.~T.,  {Flambaum} V.~V.,  {Carswell}
  R.~F.,   {Bainbridge} M.~B.,  2011, \mn@doi [\prl]
  {10.1103/PhysRevLett.107.191101}, \href
  {https://ui.adsabs.harvard.edu/abs/2011PhRvL.107s1101W} {107, 191101}

\bibitem[\protect\citeauthoryear{{Zel'dovich} \& {Novikov}}{{Zel'dovich} \&
  {Novikov}}{1966}]{YaBZel'dovich_1966}
{Zel'dovich} Y.~B.,  {Novikov} I.~D.,  1966, \mn@doi [Soviet Physics Uspekhi]
  {10.1070/PU1966v008n04ABEH002990}, \href
  {https://ui.adsabs.harvard.edu/abs/1966SvPhU...8..522Z} {8, 522}

\bibitem[\protect\citeauthoryear{de Carvalho, Rotondo, Rueda  \&
  Ruffini}{de~Carvalho et~al.}{2013}]{deCarvalho:2013rea}
de Carvalho S.~M.,  Rotondo M.,  Rueda J.~A.,   Ruffini R.,  2013, \mn@doi
  [Int. J. Mod. Phys. Conf. Ser.] {10.1103/PhysRevC.89.015801}, 23, 244

\bibitem[\protect\citeauthoryear{{de Carvalho}, {Rotondo}, {Rueda}  \&
  {Ruffini}}{{de Carvalho} et~al.}{2014}]{de2014relativistic}
{de Carvalho} S.~M.,  {Rotondo} M.,  {Rueda} J.~A.,   {Ruffini} R.,  2014,
  \mn@doi [\prc] {10.1103/PhysRevC.89.015801}, \href
  {https://ui.adsabs.harvard.edu/abs/2014PhRvC..89a5801D} {89, 015801}

\makeatother
\end{thebibliography}

\appendix
\section{Statistical analysis to minimize $\chi^2$ function}\label{Appendix}

\begin{figure}
	\centering
	\includegraphics[scale=0.5]{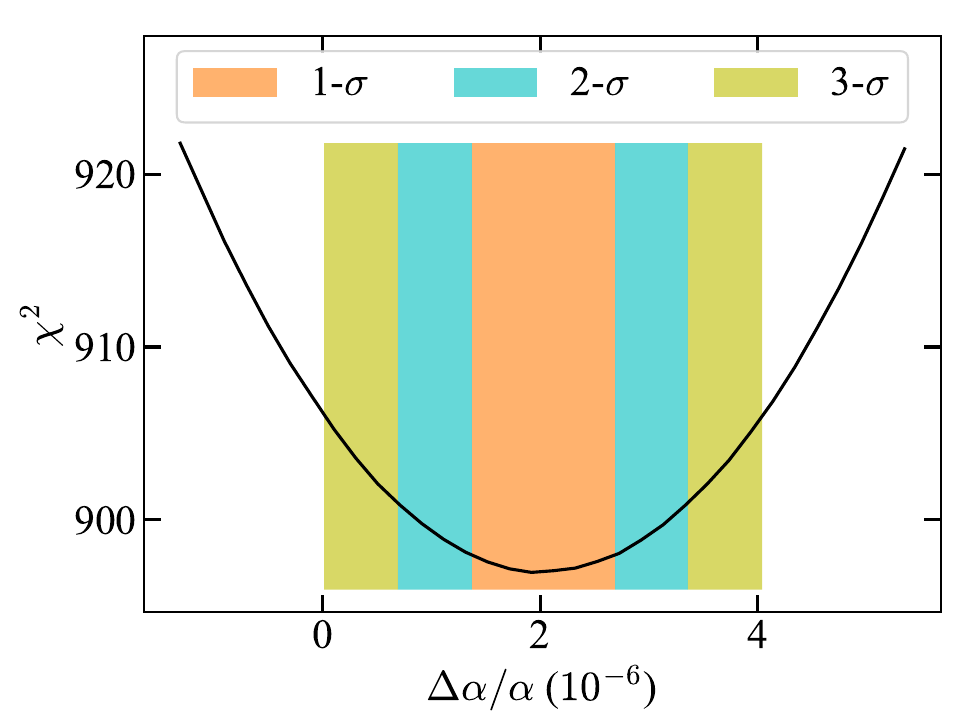}
	\caption{Variation of $\chi^2$ with respect to $\Delta{\alpha}/\alpha$ for $T=10^7\rm\,K$. Different coloured regions represent 1-$\sigma$, 2-$\sigma$, and 3-$\sigma$ confidence intervals.}
	\label{Fig: Plot}
\end{figure}

In this appendix, we discuss the statistical method used to minimize the $\chi^2$ function for the Equation~\eqref{Eq: chi square} at a specific temperature. As we have discussed in Section~\ref{Sec2}, the first step involves inserting the uncertainties of electron and proton masses in Equations~\eqref{pe} and~\eqref{rhoe} such that the uncertainties are now expressed in terms of $\Delta{\alpha}/\alpha$. We then solve the hydrostatic balance equations together with this EoS for a fixed temperature but with varied $\Delta{\alpha}/\alpha$. Using the inferred radii of the observed WDs, we calculate the $\chi^2$ function using Equation~\eqref{Eq: chi square} for each value of $\Delta{\alpha}/\alpha$. Note that we do not have the exact analytical mass--radius relation of the WDs, and hence we extract $\mathcal{R}_\text{Th}(M)$ in Equation~\eqref{Eq: chi square} using numerical interpolation. One such example of the variation of $\chi^2$ with respect to $\Delta{\alpha}/\alpha$ is shown in Figure~\ref{Fig: Plot} for a fixed temperature of $10^7\rm\,K$. We find that $\chi^2$ is minimized at $\Delta{\alpha}/\alpha = 1.932\times10^{-6}$. We also calculate different confidence intervals with this minimized value of $\chi^2$. Since we have only one free parameter ($\Delta{\alpha}/\alpha$), the 1-$\sigma$, 2-$\sigma$, and 3-$\sigma$ confidence intervals are respectively given by $\Delta{\chi^2}=1,4,9$ with $\Delta{\chi^2}$ being the change from the minimized $\chi^2$ value~\citep{press2007numerical}. Different couloured regions in Figure~\ref{Fig: Plot} depict these confidence intervals. We repeat the same procedure for other values of temperature to obtain $\Delta{\alpha}/\alpha$ corresponding to the minimized $\chi^2$ function and the confidence intervals, which are shown in Figure~\ref{Fig: constraint}(a).

\bsp	
\label{lastpage}
\end{document}